\documentstyle[12pt,fullpage,epic,eepic,float,psfig]{article}
\restylefloat{figure}
\restylefloat{table}

\title{Level Set Modeling of Transient Electromigration Grooving
\thanks{This  research  was supported by the
 Israeli Ministry of Science and Technology grant  \#9672-1-96 - 9672-3-98.}}

\author{M. Khenner$^{1}$ ~~A. Averbuch$^{1}$~~ M. Israeli$^{2}$~~M.
Nathan$^{3}$~~E. Glickman$^{3}$
\vspace{0.5cm}\\
$^{1}$Department of Computer Science \\
School of Mathematical Sciences \\
Tel Aviv University, Tel Aviv 69978, Israel\\~\\
$^{2}$ Faculty of Computer Science \\
Technion, Haifa 32000, Israel \\~\\
$^{3}$Department of Electrical Engineering-Physical Electronics \\
Faculty of Engineering \\
Tel Aviv University, Tel Aviv 69978, Israel}
\date{}

\newcommand{\Section}[1]{\setcounter{equation}{0} \section{#1}}
\newcommand{\rf}[1]{(\ref{#1})}
\newcommand{\beq}[1]{ \begin{equation}\label{#1} }
\newcommand{\eeq}{\end{equation} }

\begin{document}
\pagestyle{plain}
\maketitle
\date{}
\abstract{A numerical investigation of grain-boundary (GB) grooving
by means of the Level Set (LS) method is carried out.
GB grooving is emerging
as a key element of electromigration drift in polycrystalline
microelectronic
interconnects, as evidenced by a number of recent studies. The purpose of
the present
study is to provide an efficient numerical simulation, allowing
a parametric study of the effect
of key physical parameters (GB and surface
diffusivities,
grain size, current density, etc) on the electromigration drift velocity as well as 
on the morphology of the affected regions.
An idealized
polycrystalline interconnect
which consists of grains separated by parallel GBs
aligned normal to the average orientation of interconnect's surface
is considered.
Surface and grain-boundary diffusion are the only diffusion mechanisms
assumed.
The diffusion is driven by surface curvature gradients and by an externally
applied
electric field.
The corresponding mathematical system is an initial boundary
value problem for a two-dimensional Hamilton-Jacobi
type equation.
To solve for the electrostatic problem at a given time step, a full
model
based on the solution of Laplace's equation for the electric potential
is
employed. The resulting set of linear algebraic equations
(from the finite difference discretization of
the
equation) is solved with an effective
multigrid iterative procedure.
The details
of transient slit and ridge formation processes are presented and compared
with
theoretical predictions on steady-state
grooving \cite{GN,KGFMB1,KGFMB2}.}
\vspace{0.5cm}

\Section{Introduction}

This paper is a continuation of our work
on numerical modeling of the
formation and propagation of groove-like
defects at GBs in thin film polycrystalline  interconnects
used in microelectronics (ME).

In modern ME industry, the
reliability of ME integrated circuits has become no less
important than their performance. Some of the most vulnerable elements
of ME circuits, susceptible to several types of failures,
are the interconnects.
These are thin film metallic conductors which connect the active elements.

The defects (due to the small cross-section, high current density,
mechanical stresses and presence of GBs acting as fast diffusion
pathways) lead
to a loss (in relatively short times) of electrical and mechanical
integrity,
i.e. to line opens or shorts. For example, in the presence of a large GB
flux
($J_{gb} \approx 10^{-4}\ \mu m^2/s$) a groove can extend several
micrometers
in a few hours \cite{KGFMB1}.
Thus, GB grooving is  one of the main failure mechanisms
in advanced integrated circuits.

In the absence of an external potential field and mechanical stresses,
the GB atomic flux $J_{gb}=0$, and the corresponding groove
profile evolves via surface diffusion
under well-known
conditions of scale and temperature (the so-called Mullins problem
\cite{MULLINS57}).
In this case,
 mass transport by surface diffusion is driven only by the surface
Laplacian of curvature.
Essentially, matter flows from low-curvature regions to high-curvature regions.

In \cite{KAIN}, we presented and discussed the numerical approach
(e.g. the Level Set (LS) method) used to model GB grooving phenomena.
We also tested the LS method on two
simple, already solved, grooving problems: Mullins problem, and
that of
GB grooving by surface diffusion in a periodic array of
stationary GBs \cite{HACKNEY}. In both cases, the results obtained by
means
of the LS method are in good agreement with the theoretical predictions.
In this paper, we consider the second geometry only, as being more
realistic (see Fig. \ref{Fig1}). Due to axial symmetry at $x=0, x=L$
(where $L$
is the grain size),
we do not attempt to calculate groove branches at $x<0, x>L$.

Electric fields/currents in metallic conductors provide an
additional driving force for surface/GB mass fluxes
\cite{BLECH-HERRING}.
In the presence of an electric field, collisions between the conduction
electrons
and the metal ions lead to drift of the ions. This process is known
as electromigration (EM).

GB grooving with a GB flux in real thin film interconnects is
a complex problem.
An adequate numerical modeling technique should be capable to manage
such issues  as
GB grooving with an arbitrary EM flux, and various
ratios of GB to surface diffusivities; the latter was predicted to
critically affect groove kinetics and shape, and thus account for
various EM failure regimes (see \cite{GN,KGFMB1,KGFMB2}
and the references therein).
In cited works, analytical and semi-analytical approaches for analysing
steady-state grooving regimes were employed. However, to our knowledge,
no effort has been made to directly numerically simulate the transient
stage during GB grooving.  This is the ultimate goal of
this paper.

We do not consider mechanical stresses in GBs which,
as a matter of fact, are invariably induced by the field
\cite{BLECH-HERRING} (the approximations under
which it is reasonable to neglect the stress
are discussed in \cite{KGFMB1,KGFMB2}).
Also, under typical operational conditions of ME interconnects, lattice
transport may be neglected compared to surface/GB transport
\cite{HERRING,MULLINS63}.

Our paper proceeds as follows. In Section 2, we give details of the
physical
formulation. In Section 3, we discuss some improvements in the numerical
algorithm
and also the aspects which are due to incorporation of the electric
field/GB flux in the model
(for other algorithmic details, the interested reader should refer to
\cite{KAIN}).
Our numerical results and discussion are presented in Section 4.

\Section{Physical model}

\subsection{Driving forces for the diffusion}
\label{ele}

In the absence of an electric current, the surface diffusion is
driven by a variation in chemical potential, $\mu_s$, which causes atoms
to migrate
from high potential to low potential regions.
It was shown \cite{MULLINS57} that
\begin{equation}
\label{2.1}
\mu_s = K_s\gamma_s\Omega
\end{equation}
where $K_s$ is the surface curvature,
$\gamma_s$ is the surface tension, and $\Omega$ is the
atomic volume. Gradients of chemical potential are therefore associated
with gradients of curvature.

Let ${\bf{\tau}}$ be the tangential direction to the surface profile in
2D. Let $x,y$ be Cartesian coordinates along horizontal and vertical
boundaries of the computational box (Fig. \ref{Fig1}).
If ${\bf n}=(n_x, n_y)$ is the unit vector normal to the surface,
then the
following relations hold:
\begin{equation}
\label{2.2}
{\bf{\tau}} = (n_y, -n_x), \quad
\frac{\partial K_s}{\partial \tau}  =  {\bf \nabla} K_s \cdot
{\bf{\tau}} =
\frac{\partial K_s}{\partial x} n_y - \frac{\partial K_s}{\partial y}
n_x
\equiv K^s_\tau.
\end{equation}
The corresponding surface flux (volume crossing unit length per unit
time)
is then given by
\begin{equation}
\label{2.3}
J^{\nabla K_s}_s  = - \frac{D_s \delta_s}{k T}\frac{\partial
\mu_s}{\partial \tau} = -B \ K^s_\tau
\end{equation}
where the superscript indicates that the flux is due to the curvature
gradient,
\begin{equation}
\label{2.4}
B  =  \frac{D_s \delta_s \gamma_s \Omega}{k T}
\end{equation}
is known as Mullins constant, and $D_s, \ \delta_s, \ k$, and $T$
denote
surface diffusion coefficient, thickness of the surface diffusion layer,
Bolzmann's
constant and absolute temperature, respectively.
Note that $J^{\nabla K_s}_s$ is proportional to the first directional
derivative of the curvature.

If an electric field is present, the flux $J_s$ of matter at the curved
surface of
the conductor is driven
simultaneously by curvature gradients, and by the component $E$ of the
{\it local}
electric field along the surface. In what follows, we distinguish
between
two models which handle the electric field:

\begin{description}

\item[1. Piecewise-constant electric field]
\par\noindent

Let $C$ and $O$ denote the conductor (interconnect) material
domain and the outer (surrounding) material domain above the surface
profile, respectively
(see Fig. \ref{Fig1}).
In this model, the vector of the electric field intensity is parallel to
the GBs, ${\bf E_0} = (0,E_{0_y})$,
and $E_{0_y}$ is a step function,
\begin{equation}
E_{0_y}(x_i,y_j,t) = \left\{
\begin{array}{ll}
E_{0}^{in}=const.  & \mbox{ if grid point } (x_i,y_j) \in C \\
E_{0}^{out}=const.  & \mbox{ if grid point } (x_i,y_j) \in O. \\
\end{array}
\right.
\label{2.5}
\end{equation}
We assume that the surrounding material is less conductive than the
interconnect
material, and therefore $|E^{out}_0| < |E^{in}_0|$. In our numerical
experiments
we chose the ratio $|E^{out}_0|/|E^{in}_0| = 0.1$.
In the finite difference approach,
the discontinuous distribution of electric field intensity is 
smoothed
out across the surface profile (see the details in Section 3).
The component of the {\it local}
electric field along the surface, $E$, is then approximated by the
projection of
${\bf E_0}$ on the surface, $E={\bf E_0} \cdot {\bf \tau}$. This
approximates the
true value given by solving Laplace's equation for the potential,
subject
to the boundary conditions of
constant fields of magnitudes $E^{in}_0$ and $E^{out}_0$ in the
conductor
and surrounding material domains, respectively.

The corresponding electrically induced surface flux of matter
is given by
\begin{equation}
\label{2.6}
J^{E}_s  = - \frac{D_s \delta_s Z_s}{k T} E = - B_e \ E
\end{equation}
where the superscript indicates that the flux is due to the electric
field, and
\begin{equation}
\label{2.7}
B_e  =  \frac{D_s \delta_s Z_s}{k T}
\end{equation}
where $Z_s=z_s^* e$ is the effective charge of the ions undergoing
electromigration
in the surface layer and $e$ is the unit electronic charge;
the sign of $z_s$ is usually positive (i.e., matter flux in the
direction
of the electron flow).

\item[2. Solution of Laplace's equation for the potential]
\par \noindent

Assume that (at a given time step of overall marching algorithm) $U(x,y)$ is the electric potential within the 
(rectangular)
computational box.
$U^- (U^+)$ and $U^+ (U^-)$ are its values on the upper and lower
boundaries of the box,
and $U_n$ is the normal derivative on the boundary. $U^-$ and $U^+$ are
assumed to be
time-independent and uniform along the boundaries; 
$U^+ - U^-$ is the external voltage applied to the
interconnect.
The distribution $U(x,y)$ is governed by a static 
elliptic PDE
\begin{equation}
\label{2.8}
\frac{\partial}{\partial x}\left(k\frac{\partial U}{\partial x}\right) +
\frac{\partial}{\partial y}\left(k\frac{\partial U}{\partial y}\right) =
0,
\end{equation}
with boundary conditions $U_n=U_x=0$ on the vertical boundaries of the
box
(which in our case coincide with GBs).  Equation \rf{2.8} is derived
from
the well-posed three dimensional potential problem for the two-layer
interconnect.
The assumptions and complete derivation for the case of small aspect ratio are
presented
in \cite{AIR}. We also give some details in the Appendix.
In eq. \rf{2.8}, $k=k(x,y)$ is the specific electrical conductivity (at a given time step)
of the material which fills the computational box. To solve \rf{2.8}, a
finite
difference scheme was
developed and analysed in \cite{AIR}.
The distribution of the specific 
conductivity in the physical system under consideration is
discontinuous:
the conductivity inside the conductor material (domain $C$, Fig.
\ref{Fig1}) differs by a
finite value from that of the surrounding material (domain $O$). We
assume
\begin{equation}
\label{2.9}
k = \left\{
\begin{array}{ll}
k_{in}=const.>0  & \mbox{ if grid point } (x_i,y_j) \in C \\
k_{out}=const.>0  & \mbox{ if grid point } (x_i,y_j) \in O, \\
\end{array}
\right.
\end{equation}
i.e. $k=k(y)$ is a step function.
In our numerical experiments we chose the ratio $k_{out}/k_{in} = 0.1$.
Since the surface of the conductor evolves in time and space, then to find
the time-dependent solution $U(x,y,t)$ we need to solve the static equation \rf{2.8} 
every time step with $k$ given by \rf{2.9}. In order to be able to compute
accurately the electric field intensity (which is the derivative of $U$),
the discontinuous distribution of the specific conductivity
is smoothed out across the surface profile.
The finite difference discretization of \rf{2.8} in the computational domain leads
to a set of linear
algebraic equations with a sparse banded matrix. This set is solved with
an effective
multigrid iterative procedure \cite{AIR}.
The solution of the previous time step is used as
an initial approximation for the current step which allows fast
convergence.

After the potential is established
everywhere in the
computational domain,
the corresponding electrically induced surface flux $J^{E}_s$
is given by \rf{2.6}, where
\begin{equation}
\label{2.10}
E= -{\bf \tau} \cdot \nabla U
\end{equation}

\end{description}

To summarize the above discussion, the total
flux of matter along the surface is
\begin{equation}
\label{2.11}
J_s=J^{\nabla K}_s+J^{E}_s.
\end{equation}
Physically, equation \rf{2.11} says that atoms will diffuse in the
direction
of the electron flow if the field dominates, but toward the position
with the
large curvature if the surface energy dominates. This competition
between
the electric field and the surface energy is essential for the groove
dynamics.

The electric field results also in the diffusion of matter along GBs.
The diffusion flux along the GB, $J_{gb}$, is given by
\begin{equation}
\label{2.12}
J_{gb} = -\frac{D_{gb} \delta_{gb} Z_{gb}}{k T} E
\end{equation}
where $D_{gb},\ \delta_{gb}, \ Z_{gb}=z_{gb}^* e>0$
are the GB diffusion coefficient, thickness and effective ionic charge,
respectively,
and $E$ is the component of the
electric field along the GB.

\subsection{Boundary conditions at groove roots}

\begin{figure}[H]
\centering
\psfig{figure=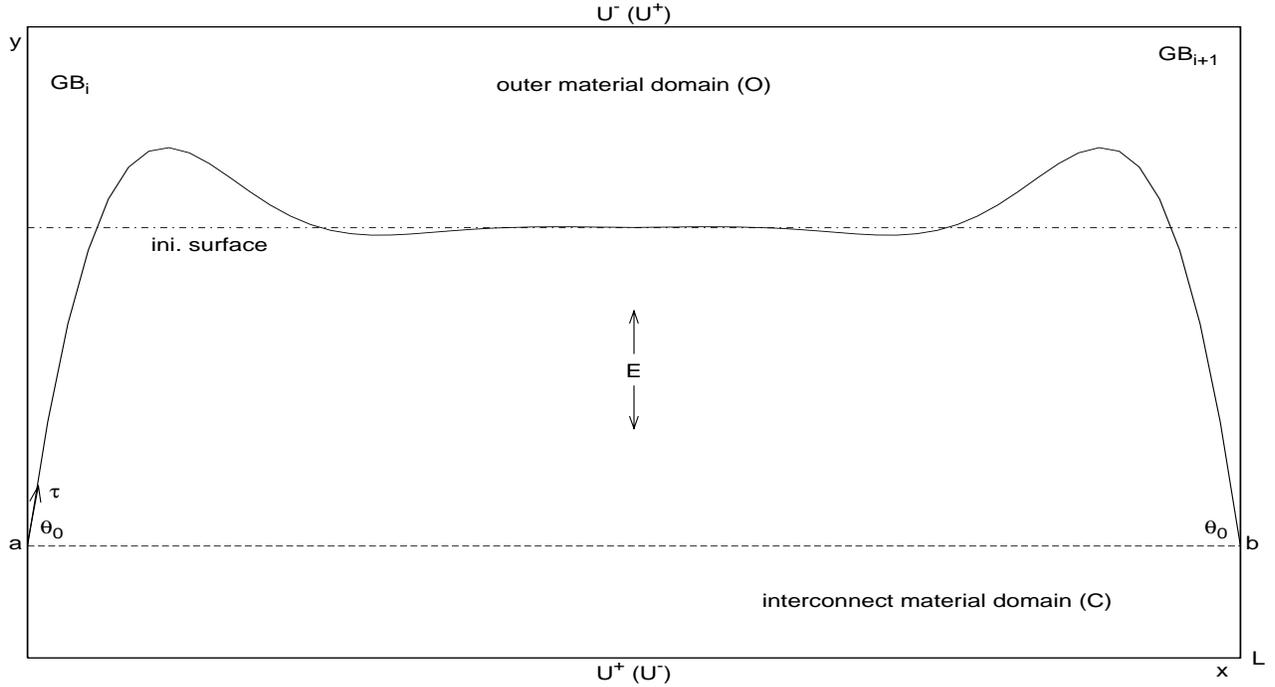,height=3.5in,width=6.5in,angle=270}
\caption{Sketch of GB grooving in a periodic array of
stationary GBs. The grain size is $L$, and groove root points are marked
as $a$
and $b$. }
\label{Fig1}
\end{figure}

The evolution of the surface is constrained by two conditions
imposed locally at groove roots $a$ and $b$ (Fig. \ref{Fig1}):

\begin{description}

\item[1. Equilibrium angles]
\par \noindent

The boundary condition is dictated by the local
equilibrium between the surface tension, $\gamma_s$, and the GB tension,
$\gamma_{gb}$. In the symmetric case of a GB ($x=0$) normal to an
original ($y=const.$) flat surface, the angle of inclination of the
right branch
of the surface at the groove root with respect to the $x$ axis is
\cite{MULLINS57}
\begin{equation}
\label{2.13}
\theta_0 = sin^{-1}(\gamma_{gb}/2\gamma_s)=const.
\end{equation}
The rapid establishment of the equilibrium angles between the GBs
and the surface by atomic migration in the vicinity of the intersections
develops some curvature gradients at the adjacent surface,
and thus
induces surface diffusion fluxes, $J^{\nabla K}_s$, along the groove
walls.
The directions of the fluxes depend on the sign of the respective
surface curvature
gradients at the groove groots.

\item[2. Continuity of electrically induced fluxes]
\par \noindent

The boundary conditions read
\begin{equation}
\label{2.14}
J_{gb}=2 J^{E}_s,
\end{equation}
since both branches of the groove act as  sinks or sources
of matter.

\end{description}

\Section{The numerical procedure}

The Level Set method is used to ``capture" the evolution of the
conductor
surface.
The method was introduced by Osher and Sethian \cite{SETYAN-BOOK2}
and was further
developed during the last several years.
The method enables to capture drastic changes in the
shape of the curves (surfaces or interfaces) and even topology changes.

The basic idea of the method consists of embedding the curve $y(x,t)$
into a higher dimensional space. As a matter of fact, we consider the
evolution of a two-dimensional field $\phi(x,y,t)$ such that its zero
level set, $\phi(x,y,t)=0$, coincides with the curve of interest,
$y(x,t)$, at any time $t$. The level set function $\phi(x,y,t)$
can be
interpreted as a signed distance from the curve $y(x,t)$, which moves in
the
direction normal to itself.

The evolution of $\phi(x,y,t)$ is described by an Hamilton-Jacobi
type equation. A remarkable trait of the method is that the function
$\phi(x,y,t)$ remains smooth, while the level surface $\phi = 0$ may
change
topology, break, merge, and form sharp corners as $\phi$ evolves. Thus,
it is possible to perform numerical simulations on a discrete grid
in the spatial domain, and substitute  a finite difference
approximations
for the spatial and temporal derivatives in time and space.
Another nice feature of the method is that the explicit location of the
interface
needs not to be known in the computational process; all the necessary
information
is extracted from the level set function.

The evolution equation has the form
\begin{equation}
\label{3.1}
\phi_t + F |\nabla \phi| =  0 \quad \mbox{given}\ \ \phi(x,t=0).
\end{equation}
%
The normal velocity, $F$, is considered to be a function of spatial
derivatives of $\phi(x,y,t)$. In many applications $F$ is a function of
the curvature, $K_s$, and its spatial derivatives. The curvature $K_s$
may
be computed
via the level set function $\phi$ as follows:
\begin{equation}
\label{3.2}
K_s  =  \nabla \cdot {\bf n}, \quad {\bf n}  =
\frac{ \nabla \phi}{|\nabla
\phi|}=\left(\frac{\phi_x}{\left(\phi_x^2+\phi_y^2\right)^{1/2}},
\frac{ \phi_y}{\left(\phi_x^2+\phi_y^2\right)^{1/2}}\right).
\end{equation}
Here ${\bf n}$ is a \lq \lq normal vector", and it coincides with the
(previously introduced)
unit
normal to the surface, $y(x,t)$, on the zero level set $\phi=0$.
Formulas \rf{3.2} can be combined as follows
\begin{equation}
\label{3.3}
K_s  =  \nabla \cdot \frac{ \nabla \phi}{|\nabla \phi|} =
\frac{\phi_{xx} \phi_y^2 - 2 \phi_{x} \phi_y \phi_{xy} + \phi_{yy}
\phi_x^2}
{ \left( \phi_{x}^2 + \phi_y^2 \right)^{3/2}},
\end{equation}
and the sign of $K_s$ is chosen such that a sphere has a positive mean
curvature
equal to its radius. In the case of surface diffusion in 2D,
\begin{equation}
\label{3.4}
F  =  \frac{\partial J_s}{\partial \tau} =
\frac{\partial J^{\nabla K}_s}{\partial \tau}+\frac{\partial
J^E_s}{\partial \tau}
\end{equation}
where $J_s$ is given by \rf{2.11}.

The difficulties in the numerical solution of \rf{3.1} in our case are
due to the fact that, as could be noted from \rf{2.3}, \rf{3.3},
\rf{3.4},
the first term in $F$ contains
space derivatives of order 4 of the level set function.
Therefore, the evolution equation \rf{3.1} is highly sensitive to
errors. Besides, this fourth derivative
term leads to
schemes with very small time steps.

In \cite{KAIN}, we presented the computational algorithm which solves
the problem of GB grooving by surface diffusion in the absence
of electromigration.
This could be viewed as the limiting case of the problem
which is under consideration in this paper, corresponding to
the situation where electrically induced surface and GBs fluxes
$J^E_s$ and $J_{gb}$ vanish. The normal velocity function \rf{3.4}
in the latter case contains only the first term.
The basic features of the algorithm are:

\begin{itemize}

\item the use of a uniform static grid in both space directions

\item the use of a standard
      second order-accurate finite difference scheme in space

\item the approximation of spatial derivatives (in normal direction) on
the boundaries
      of the computational box by second order one-sided
differences

\item time marching is done by a second-order Total Variation
Diminishing
      (TVD) Runge--Kutta procedure \cite{OSHER-SHU,SHU-OSHER}

\item the use of second-order Essentially Non-Oscillatory (ENO) scheme
      \cite{ZCMO} to approximate the gradient function in \rf{3.1}

%
%

\item the use of ``reinitialization" \cite{SSO} every time step to keep
the
      level set function $\phi$ a signed distance function

\end{itemize}

\begin{figure}[H]
\centering
\psfig{figure=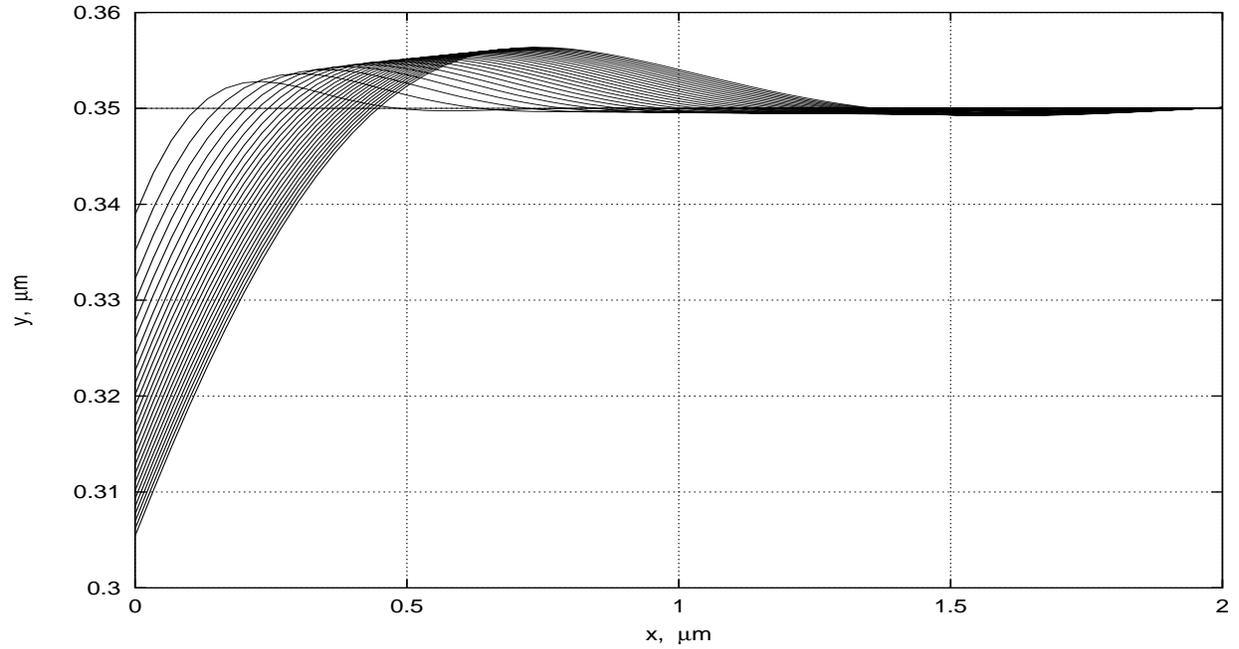,height=3.5in,width=6.5in,angle=270}
\caption{Groove development and propagation along the GB of an infinite
bicrystal
(Mullins problem). Grain size $L=2\ \mu m$.
For copper interconnects at $T=600\ K\ (327\ ^{o}C)$, $B = 9.2\times
10^{-33}\ m^4/s$.
The angle at the
groove root $\theta_0=\pi/22\ (\tan{\theta_0} \approx 0.144)$. }
\label{Fig2}
\end{figure}

\begin{figure}[H]
\centering
\psfig{figure=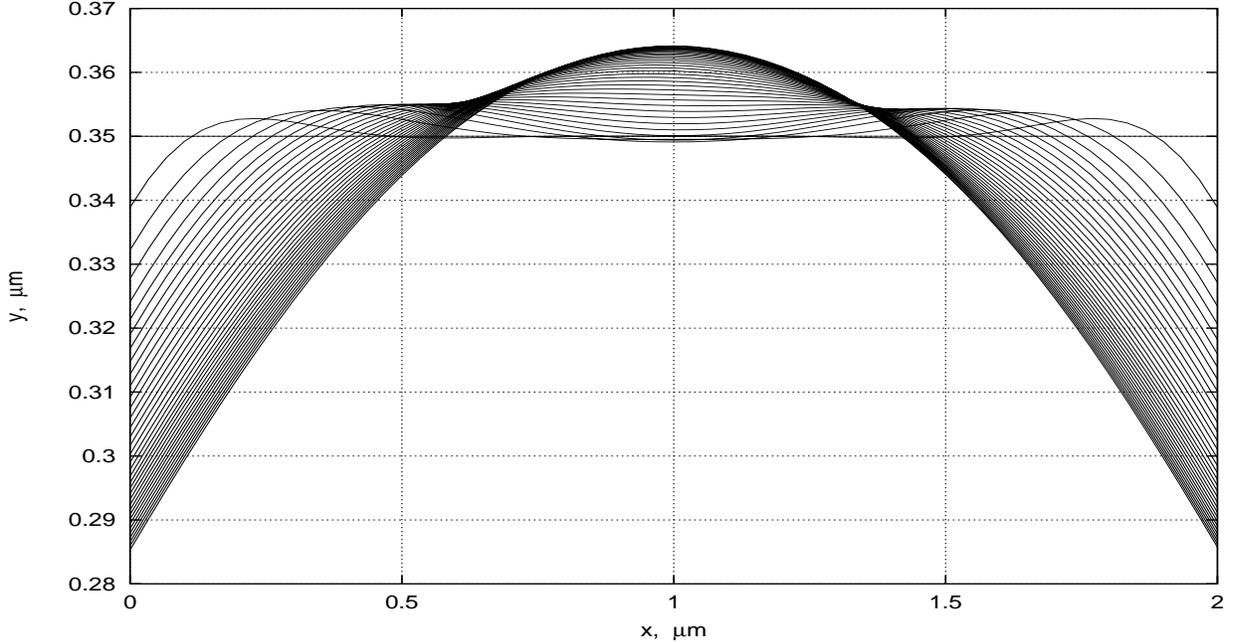,height=3.5in,width=6.5in,angle=270}
\caption{GB grooving in a periodic array of
stationary GBs. Parameters are the same as in Fig. \ref{Fig2}.}
\label{Fig3}
\end{figure}

\noindent
The solution of \rf{3.1} (in Mullins case of an infinite bicrystal with
a single GB)
subject to the conditions of a
constant angle of surface inclination and zero surface flux
$J^{\nabla K}_s$ at the groove root $a$, is then a self-similar surface
profile, whose linear dimensions are proportional to $(Bt)^{1/4}$,
$B$ given by \rf{2.4} (Fig. \ref{Fig2}). If the dimensions of
the crystal are finite, grooves develop at each GB;
grooving stops when, at sufficiently long times, identical circular
arcs develop connecting adjacent GBs (Fig. \ref{Fig3}).
The parameters chosen for the runs are typical for copper interconnects
at temperatures relevant to experiments ($T=600\ K$) \cite{GN,MILLINS}:
$\Omega=1.18\times 10^{-29}\ m^3,\ D_s=3.3\times 10^{-14}\ m^2/s,\
\gamma_s=1.7\ J/m^2,\
\delta_s=3.5\times 10^{-10}\ m,\ kT=8.28\times 10^{-21}\ J$.
It is worth noting that our numerical treatment of GB grooving
is not constrained by the assumption of small equilibrium angles
(``small slope approximation"),
in contrast to the analytical approaches of the pioneer works
\cite{HACKNEY,MULLINS57}.

In \cite{KAIN}, special attention was given to the treatment of
constant-angle and zero-flux conditions at the groove roots
within the framework of the Level Set method.
Two methods were developed, the first based on interface reconstruction
every time step, with subsequent correction of the angles followed
with reinitialization, and the second based on the extension of the
$\phi$-field beyond the GBs using the expansion in Taylor series
up to second order.
Both methods were successfully used in calculations, but we
observed that sometimes both procedures resulted in a loss of accuracy.
In this paper, we propose a new, robust and highly accurate
procedure to keep the equilibrium angle \rf{2.13} constant at the
intersections of the surface with the GBs.

Consider equations \rf{2.2}, \rf{3.2} and the zero level line of $\phi$
(conductor's surface) passing through the groove root point $a$
in Fig. \ref{Fig1} (a similar analysis could be performed for groove
root point $b$).
Since the tangential vector to the zero level line at $a$ (as well as to
other
level lines at $x=0$ if $\phi$ is kept a signed distance function), is
${\bf \tau} = (\cos{\theta_0}, \sin{\theta_0})$,
then \rf{2.2}, \rf{3.2} imply
\begin{equation}
\label{3.5}
n_x=-\sin{\theta_0}=\frac{\phi_x}{\left(\phi_x^2+\phi_y^2\right)^{1/2}},
\quad
n_y=\cos{\theta_0}=\frac{\phi_y}{\left(\phi_x^2+\phi_y^2\right)^{1/2}}.
\end{equation}
Dividing the first equation in \rf{3.5} by the second one gives
\begin{equation}
\label{3.6}
\phi_x=-\phi_y\ \tan{\theta_0}.
\end{equation}
As pointed out above, we approximate the spatial derivatives (in normal
direction)
of $\phi$ at the boundaries of the computational domain by second order
one-sided
differences. Therefore, equation \rf{3.6} written on the left boundary
$x=0$ takes the form
\begin{equation}
\label{3.7}
\frac{-3\phi_{0,j}+4\phi_{1,j}-\phi_{2,j}}{2\Delta x}=
-\frac{\phi_{0,j+1}-\phi_{0,j-1}}{2\Delta y} \tan{\theta_0}, \quad
j=0,...m-1
\end{equation}
where $m$ is the number of grid points in the vertical direction, and
$\Delta x$ and
$\Delta y$ are grid spacings in the horizontal and vertical directions,
respectively. Rearranging the terms
in \rf{3.7} gives the following set of nonhomogeneous
linear algebraic equations with a tridiagonal matrix for unknowns
$\phi_{0,j},\ j=0,...m-1$:
\begin{equation}
\label{3.8}
-\frac{\tan{\theta_0}}{2\Delta y}\phi_{0,j-1}-\frac{3}{2\Delta
x}\phi_{0,j}+
\frac{\tan{\theta_0}}{2\Delta y}\phi_{0,j+1}=
-\frac{4\phi_{1,j}-\phi_{2,j}}{2\Delta x}.
\end{equation}
The solution to this (and to the similar set at $x=L$) is easily and
accurately found at the beginning of every stage of a Runge--Kutta time
marching,
thus providing the field $\phi$ which incorporates the correct
equilibrium
angles at the groove roots.

The described procedure allows us to have a straight horizontal line
\begin{equation}
\label{3.9}
y(x,0)=const.
\end{equation}
where $const.>0$ gives the initial height of the material domain, as
initial condition
for LS simulations.
Note that initial condition \rf{3.9} does not match the boundary
condition \rf{2.13}. This implies that a singularity exists at $x=0,
x=L$ at $t=0$.
This singularity does not present a barrier in solving the system
numerically
when we select an appropriate numerical scheme. Physically, the
equilibrium angle
is formed instantaneously compared with the time needed for the
evolution of
the surface. We are not concerned with the details of this instance.
We could use an initial surface which is consistent with the boundary
conditions
(as done in \cite{KAIN} where Mullins profile
served as an initial condition, and we followed the evolution of this
profile in time).
However, the choice of initial condition \rf{3.9} is more physical.

To close this section, we present the details of the calculation of the
normal velocity
function \rf{3.4}:

\begin{enumerate}

\item Calculate curvature induced flux $J^{\nabla K_s}_s$ from \rf{2.3}.
      It is nonzero even at the first time step, since the equilibrium
angles \rf{2.13}
      are formed instantly

\item Calculate the first term in \rf{3.4} by applying the formula
\beq{3.10}
\frac{\partial J^{\nabla K}_s}{\partial \tau} =
\nabla\left[\nabla J^{\nabla K}_s \cdot {\bf \tau} \right]\cdot {\bf
\tau} =
\eeq
$$
B\left[\frac{-K_{xx} \phi_y^2 + 2 K_{xy} \phi_x \phi_{y} - K_{yy}
\phi_x^2}
{ \phi_{x}^2 + \phi_y^2}+
K\left(K_y(n_x+n_y)-K_x(n_y-n_x)\right)\right]+KJ^{\nabla K_s}_s
$$

\item Solve the electrical problem, find electrically induced surface
flux
      $J^E_s$ from \rf{2.6}. As pointed out above, the discontinuous
distributions
      of electrical  quantities are smoothed across the surface profile
- by a hyperbolic
      tangent law:
\beq{3.11}
r=\frac{r_{out}+r_{in}}{2}+\frac{r_{out}-r_{in}}{2}\tanh{\beta\phi(x,y)}
\eeq
where $\beta$ is a large constant adjusting parameter, and
$r$ is either the electric intensity $E(x,y)$ (first electrical model),
or the
specific conductivity $k(x,y)$ (second electrical model)

\item Given values of the electrical intensity $E$ along the GBs,
calculate
      electrically induced GB fluxes, $J_{gb}$, from \rf{2.12}.
      Applying boundary condition \rf{2.14}, calculate corrected values
      of $J^E_s$ along grid lines $x=0,\ x=L$

\item Calculate the second term in \rf{3.4} by applying the formula
\rf{2.2}
      where $J^E_s$ replaces $K_s$.

\end{enumerate}

\Section{Numerical results and discussion}

Several comments should be made before we present the results
of the numerical simulations.

\begin{itemize}

\item Due to the large number of material parameters
involved we concentrate on the influence of the one which
was predicted to greatly affect the grooving process, i.e.
the ratio of the GB to surface diffusivity, $r_d=D_{gb}/D_s$ \cite{GN}.
The parameter set we choose for the simulations corresponds
to copper, $Cu$, at temperatures about 600 K. It should be noted
that ({\bf i}) the
experimentally measured values of diffusivities could vary,
according to different sources, by up to 3 orders of magnitude, and ({\bf ii})
$D_{s}$ can be smaller than $D_{gb}$,
due to, for example, surface contamination, thus giving
$r_d > 1$. Accordingly,
we fix the value of $D_s$ and vary $D_{gb}$ in a wide range,
thus varying the GB flux $J_{gb}$ \rf{2.12}.

\item We study the advancing (elongating) grooves characterized
by the positive values of $J_{gb}$ (matter flows out of the
groove cavity and into the GB). In this case the electric field
intensity vector is directed upwards (see Fig. \ref{Fig1}), the positive potential
being prescribed on the lower boundary of the computational
box and the negative potential on the upper one. Reversing  the
direction of the field produces receding grooves or ``ridges",
Fig. \ref{Fig8},
characterized by a negative GB flux (matter flows out of the GB  into the
groove cavity). The advancing grooves are of more practical interest,
as explained in the Introduction.

\item Most of our results are obtained with electrical
model 2 (see section \ref{ele}), based on the solution of Laplace's
equation for the
potential. As expected, this model produced more accurate
results than the approximate model 1 based on the piecewise-constant
electrical field. However, model 1 \footnote{Or even more simplified
model of the constant field throughout the entire domain $C+O$
(Fig. \ref{Fig1}) used in most,
if not
all studies of electromigration
(see \cite{KGFMB2} for example)} proved to be rather
useful in our simulations, since it produced qualitatively
good results in greatly reduced (comparable to the model 2)
computational times; for very fine
grids ($80\times80$ resolution) the speedup achieved
is by as much as a factor of 6.

\end{itemize}

\begin{figure}[H]
\centering
\psfig{figure=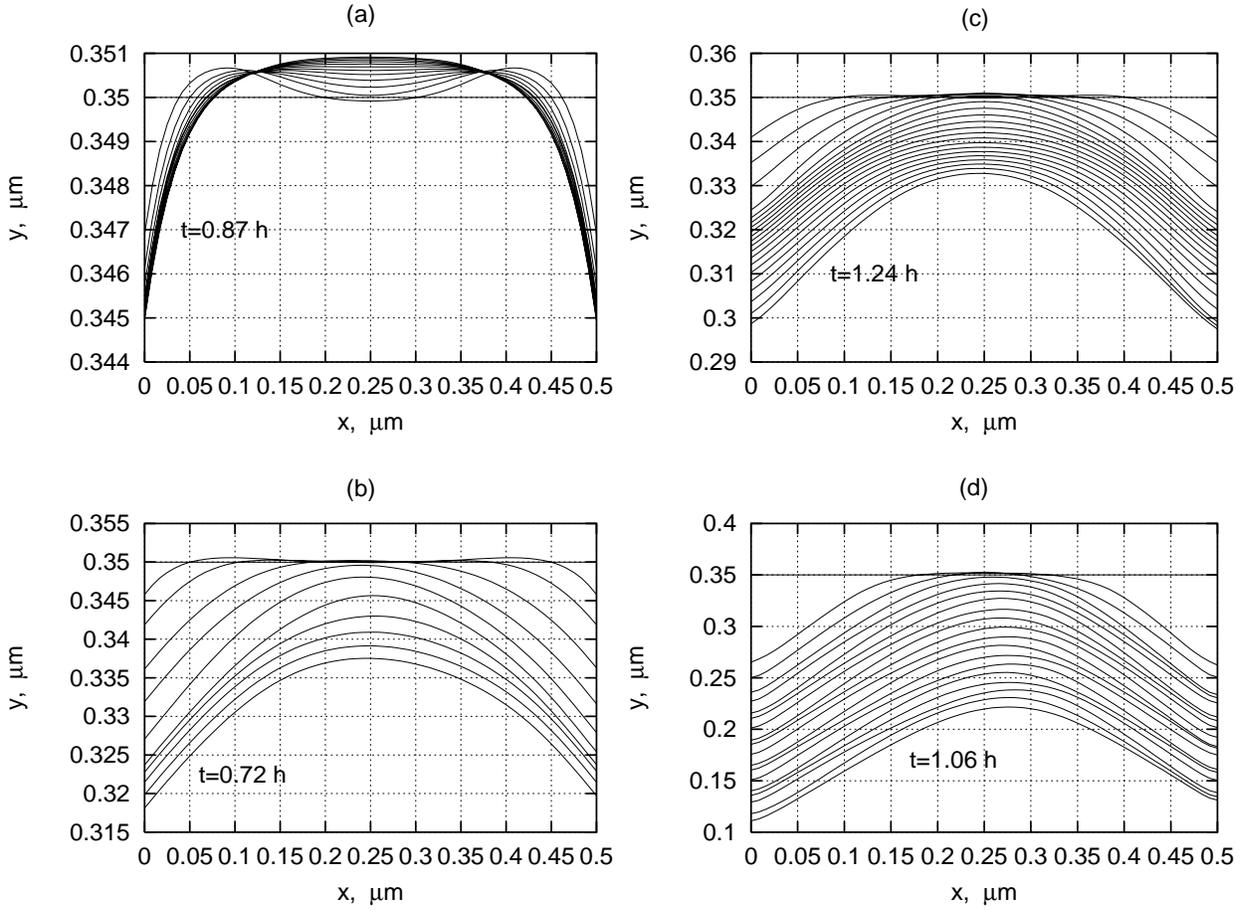,height=4.7in,width=6.5in,angle=270}
\caption{GB grooving by surface/GB diffusion driven by the surface
curvature gradients and the electromigration.
(a): $r_d=0.224$, (b): $r_d=0.336$, (c): $r_d=0.561$, (d): $r_d=22.424$.
The surface profiles are dumped every 5000 time
steps.
The time labels correspond to the (physical) time
at which the last profile is dumped.}
\label{Fig4}
\end{figure}

Fig. \ref{Fig4} (a)-(d) shows the space/time evolution of the initially
flat
surface of the conductor for different values of $r_d$.
The parameters are as in Figs. \ref{Fig2}, \ref{Fig3} and
$U^+=-U^-=5.0\times
10^{-3}\ V,\ k_{in}=10^8\ (\Omega m)^{-1},\ k_{out}=10^7\ (\Omega
m)^{-1},\
\delta_{gb}=\delta_s=3.5\times 10^{-10}\ m,\
z_s^*=z_{gb}^*=5$.
The surface profiles are dumped every 5000 time
steps,
the dimensions of the computational box are $0.5\ \mu m\times 0.5\ \mu
m$, and
the grid has a $60\times60$ resolution.

\noindent
In case $r_d$ is much less than one (Fig. \ref{Fig4} (a)), i.e.
the GB flux
is relatively small, and we observe that the evolution of the surface
is similar to the one shown in Fig. \ref{Fig3}. It slows down
in time, providing to be not very dangerous in the sense of failure
of the conductor. The evolution proceeds
faster as $r_d$ increases (Fig. \ref{Fig4} (b)). After the GB grooves
merge and form a single profile, this profile starts to advance slowly
(note that the surface curvatures at the groove roots are
still positive, at least during the time of the observation).
Yet larger values of $r_d$ (Fig. \ref{Fig4} (c)) result in changes of the morphology of
the
surface profile in the near-groove-tip regions. The latter means that
the sign of the surface curvatures at the groove tips changes
from being positive to a negative one in a realively short time
after the evolution starts,
indicating the surface tendency to form so-called slits.
In the case of Fig. \ref{Fig4} (c) this
transition takes place at the time step $n^*,\ 15000 < n^* > 20000$;
see also Figures \ref{Fig6}, \ref{Fig7}.
No qualitative change in the surface shapes is observed as $r_d$ is further
increased - up to the limit where the numerical method is applicable
(note the significant losses of accuracy in Fig. \ref{Fig4} (d), which
corresponds to $r_d=22.424$).
Fig. \ref{Fig4} (d) differs from Fig. \ref{Fig4} (c) only in
the increased velocity of the surface's advance and in a more rapid transition
from positive to negative curvature at the groove roots (the decreased $n^*$).
The evolution regime  shown in Fig. \ref{Fig4} (c), (d) is known
as the $A$-regime \cite{GN,KGFMB1}.

\begin{figure}[h]
\centering
\psfig{figure=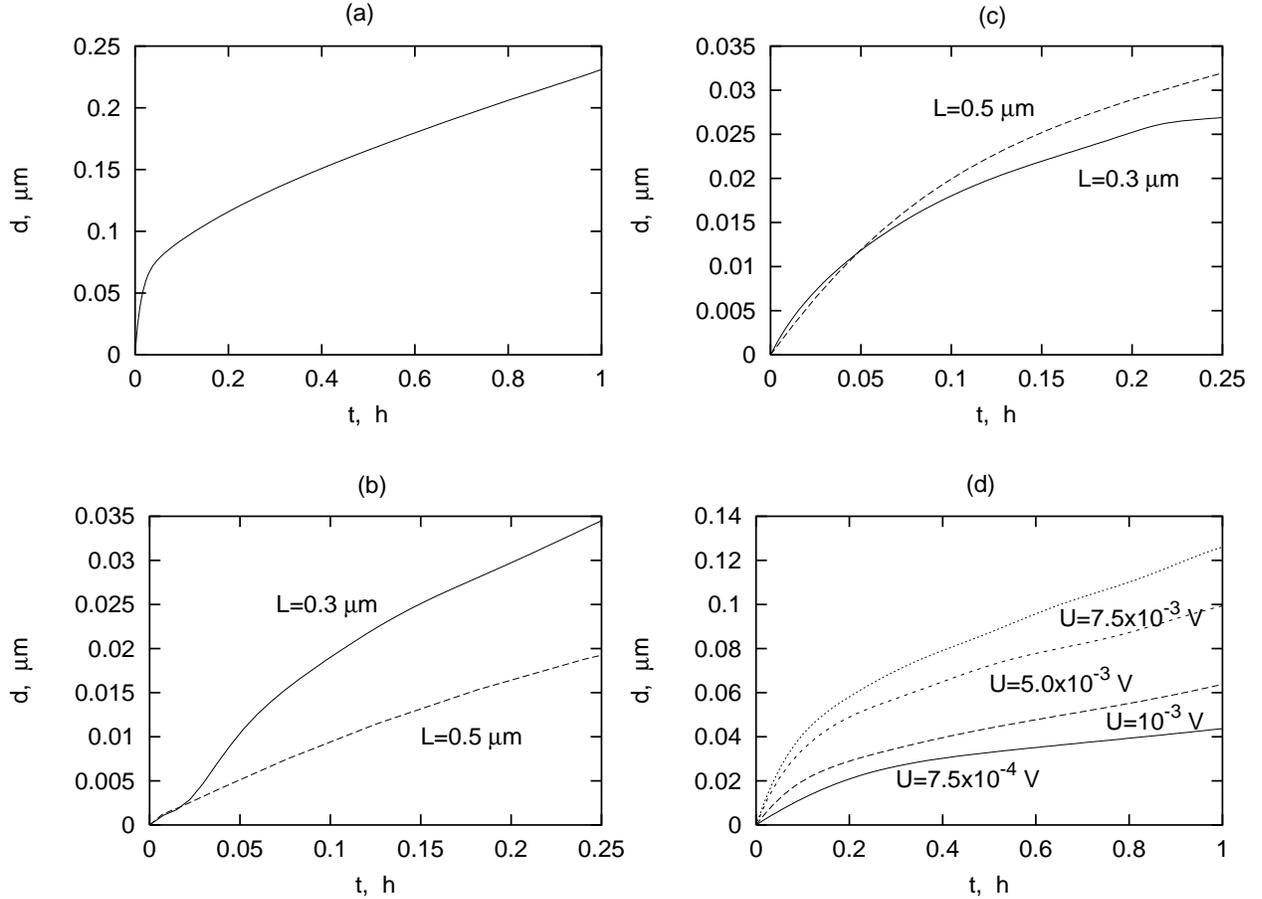,height=4.7in,width=6.5in,angle=270}
\caption{Distance $d$ traveled by the
groove tips as a function of time.
(a): $r_d=22.424$, refer
to Fig. \ref{Fig4} (d);
(b): $r_d=2.424,\ U=1.0\times 10^{-4}\ V$;
(c): $r_d=2.424,\ U=1.0\times 10^{-3}\ V$;
(d): $r_d=2.424,\ L=0.5\ \mu m$.}
\label{Fig5}
\end{figure}

\begin{figure}[H]
\centering
\psfig{figure=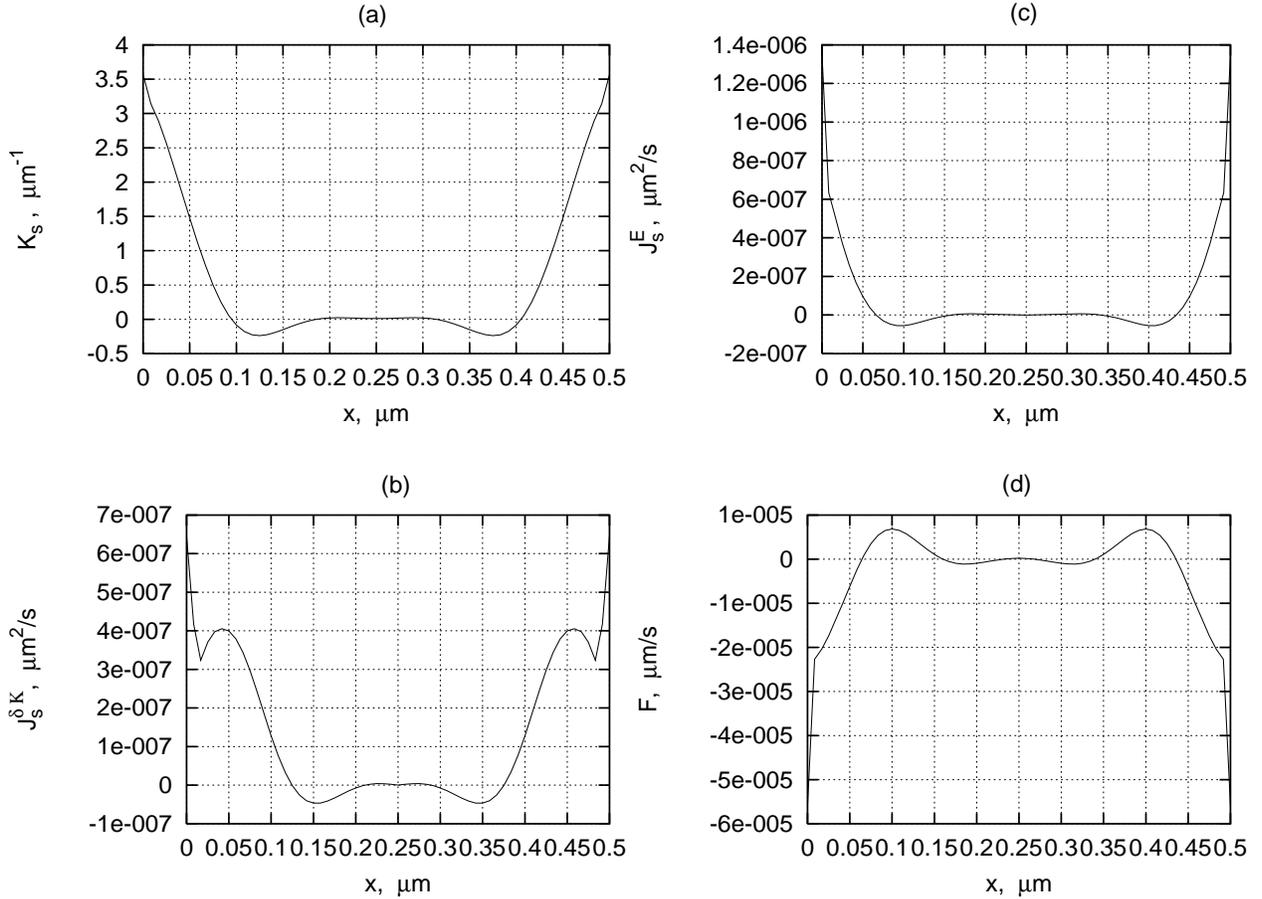,height=4.7in,width=6.5in,angle=270}
\caption{Plots of the surface's curvature (a), diffusion fluxes (b), (c)
and
the normal velocity function (d). $r_d=0.561$, refer to Fig. \ref{Fig4}
(c).
The corresponding time step is 1000.}
\label{Fig6}
\end{figure}

\begin{figure}[H]
\centering
\psfig{figure=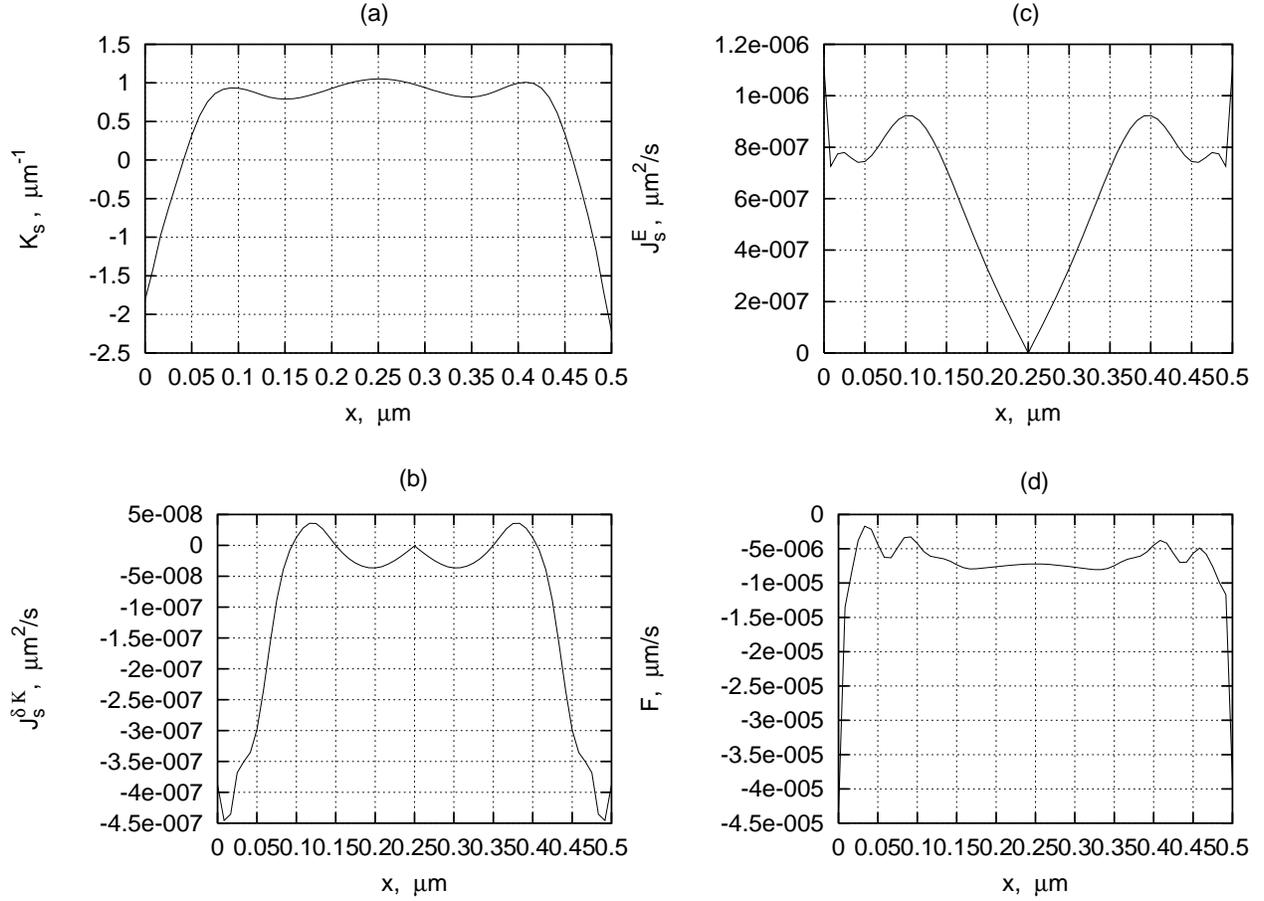,height=4.7in,width=6.5in,angle=270}
\caption{Same as Fig. \ref{Fig6}, but the data correspond to the
time step 35000 at Fig. \ref{Fig4} (c).}
\label{Fig7}
\end{figure}

\begin{figure}[H]
\centering
\psfig{figure=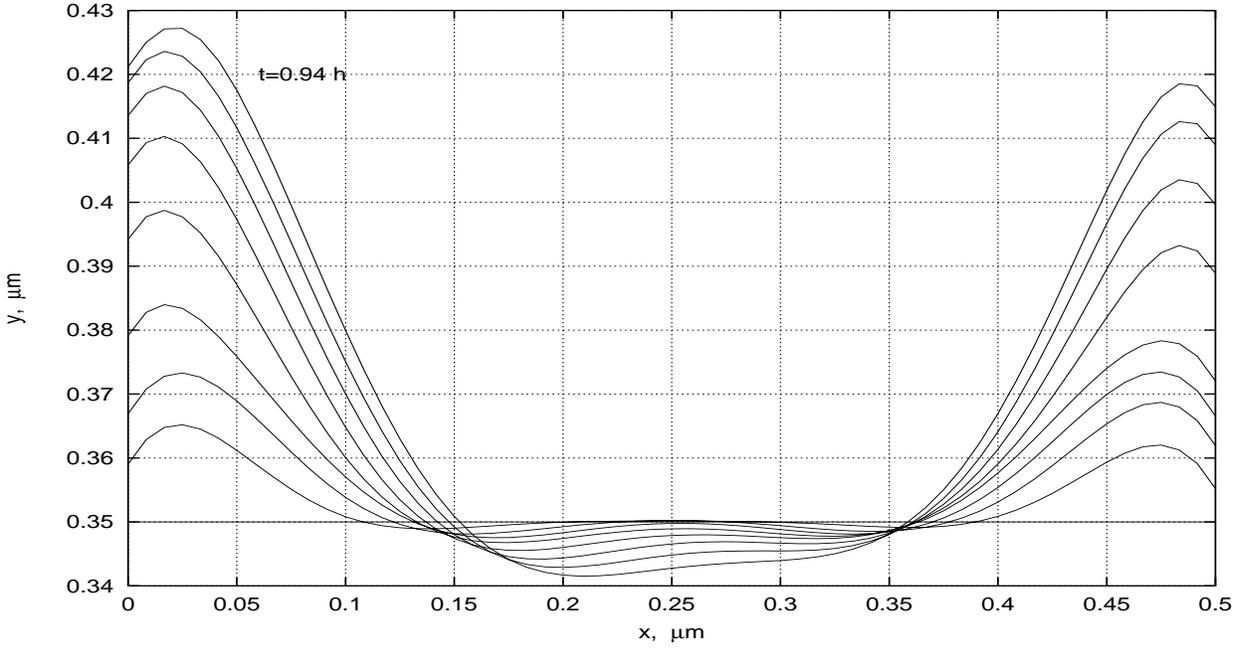,height=3.5in,width=6.5in,angle=270}
\caption{The ridges formed if the matter flows into the groove tips
($r_d=1.121$, the electric field intensity vector is directed from top
to bottom).
The physical parameters are as in Fig. \ref{Fig4}.}
\label{Fig8}
\end{figure}

In Fig. \ref{Fig5} (a) - (d) we plot the distance $d$ traveled by the
groove tip as a function of time. Fig. \ref{Fig5} (a) corresponds to the
case of
Fig. \ref{Fig4} (d) ($r_d=22.424,\ L=0.5\ \mu m,\ U=U^+=U^-=5.0\times
10^{-3}\ V$).
One sees that the steady-state velocity of the
surface advance is attained rather rapidly, within approximately
6 min. Also note that the groove tip traveled (in only
1 hour) a distance
which is a little less than half the grain size.
In Fig. \ref{Fig5} (b), (c) we investigate the influence of the grain
size, $L$.
When $r_d$ is small and the applied voltage is small too (Fig.
\ref{Fig5} (b),
$r_d=2.424,\ U=1.0\times 10^{-4}\ V$), the evolution is driven mostly by
the surface flux $J^{\nabla K}_s$. Then, smaller sizes of the grains
result in
larger velocities of the groove tip. This is because the curvature of
the surface
increases as the grain size decreases, resulting in the increase of
$J^{\nabla K}_s$.
The example of such a transitive grooving regime  (from classical Mullins regime, Fig.
\ref{Fig4} (a)
to $A$-regime, Fig. \ref{Fig4} (c), (d))
 is presented in Fig. \ref{Fig4} (b).
If electrically induced fluxes $J^E_s$ and $J_{gb}$ are dominant
($A$-regime, Fig. \ref{Fig5} (c),
$r_d=2.424,\ U=1.0\times 10^{-3}\ V$) then, in contrast with the previous
case,
larger grain sizes result in larger groove tip velocities, as
predicted in \cite{KGFMB1}. The dependence of the
groove tip velocity on the applied voltage is illustrated by Fig.
\ref{Fig5} (d)
($r_d=2.424,\ L=0.5\ \mu m$). GB grooving proceeds faster in strong
electric fields
due to the amplification of the electromigration and associated
diffusion fluxes
$J^E_s$ and $J_{gb}$.

For completeness, in Fig. \ref{Fig6} (a)-(d) and in Fig. \ref{Fig7}
(a)-(d)
we present plots of the surface curvature, the $J^{\nabla K}_s$ and
$J^E_s$ diffusion fluxes, and the normal velocity function $F$ for the
case of Fig. \ref{Fig4} (c). The data in Fig. \ref{Fig6} correspond to
time step
1000 (transient stage), while the data in
Fig. \ref{Fig7} correspond to time step
35000 (steady-state stage). Grooves develop faster
during the transient stage (compare Fig. \ref{Fig6} (d) and Fig.
\ref{Fig7} (d),
see also Fig. \ref{Fig5}),
when the curvature of the surface in the near-groove tips regions is
positive
and both $J^{\nabla K}_s>0$ and $J^E_s>0$ fluxes tend to elongate the
grooves,
providing
the flow of matter out of the groove tips.
As the steady-state approaches, the curvature of the surface in the
near-groove
tips regions becomes negative
(Fig. \ref{Fig7} (a)) and the $J^E_s$ flux is still into the GB
but the flux $J^{\nabla K}_s$ changes the direction and slows the
evolution down
(Fig. \ref{Fig7} (b), (c)).

\begin{figure}[H]
\centering
\psfig{figure=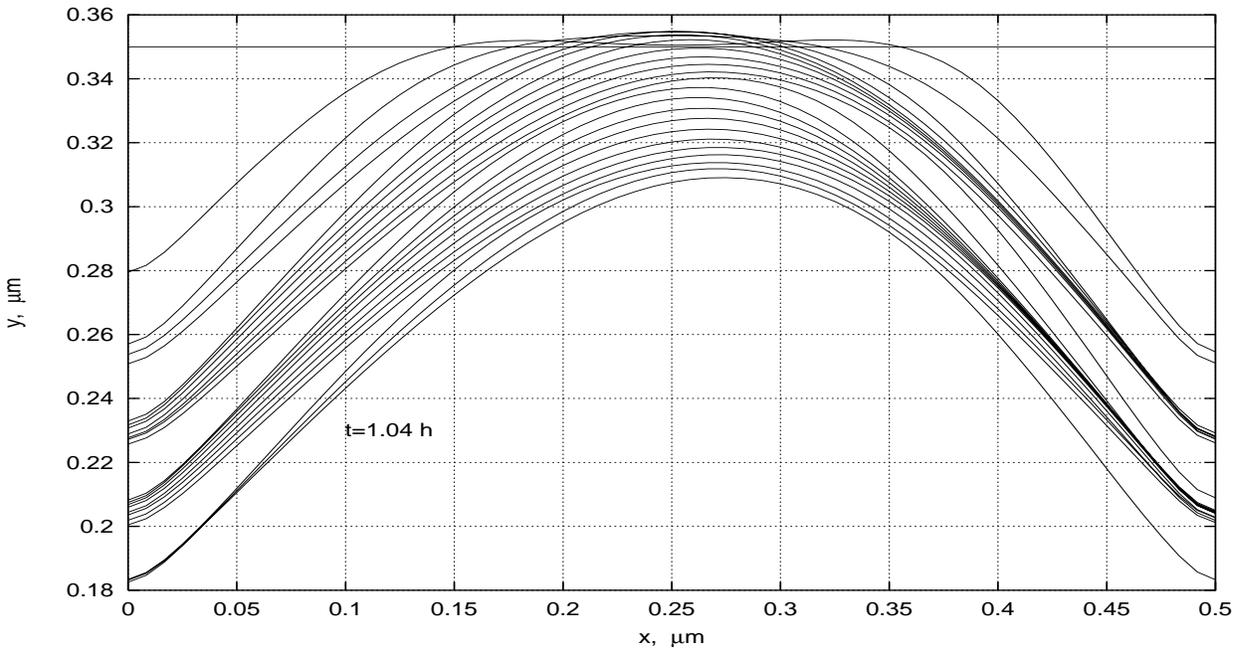,height=3.5in,width=6.5in,angle=270}
\caption{GB grooving, $r_d=22.424$ (compare to Fig. \ref{Fig4} (d)).
The electrical model 1 was used.}
\label{Fig9}
\end{figure}

Fig. \ref{Fig9} shows the $A$-regime of GB grooving obtained with the use of the
electrical model 1 (see section \ref{ele}; to be compared with the Fig.
\ref{Fig4} (d)).
The computations are less accurate if this model is 
employed, resulting in highly 
asymmetric surface profiles. 
However, the dynamics of surface 
evolution could be 
predicted from these 
simulations and we made a heavy use of the electrical model 1 
for trial 
numerical experiments. 
It is worth to note that the run time to obtain
Fig. \ref{Fig9} is 2.2 hours on SGI workstation with 194 MHz IP25 processor,
compared to 8.1 hours for Fig. \ref{Fig4} (d). 

\bigskip
\Section{Conclusions}
The Level Set method was used to model the GB grooving
by surface/GB diffusion in an idealized polycrystalline
interconnect.
The diffusion is driven by surface curvature gradients and external
applied
electric field.
The results demonstrate the high potential of the LS method for the
simulation
of complex failure phenomena in microelectronic interconnects.
The plans for future research are:

\begin{itemize}

\item to obtain more physical results with the current
version of the code and compare them to experimental ones

\item to improve our numerical procedure to make possible
the simulation of the propagation of slits ($B$-regime,
\cite{GN,KGFMB1,KGFMB2}).
The latter are
characterized by (almost) straight vertical walls
\footnote{The flux $J^{\nabla K}_s$ is zero along these walls.}
and negative curvatures at the slit tips. The slits are formed at high
values
of $r_d$,
and are supposed to propagate in a {\em local} steady state, leaving the
rest of the surface behind.
Physically, the surface of the conductor cannot accomodate a very large
GB flux;
the groove
tips then become diffusively detached from the remaining surface.
At the moment our numerical procedure does not
allow to fully trace the evolution of the slits and the surface left
behind. In our opinion, a locally refined grid is needed to provide high
accuracy in the
near-slit-tip regions; however, the adaptation of the LS method
to such grids may be not straightforward. In our simulations
(which make use of an uniform grid),
the instability steps in shortly.

\item to incorporate mechanical stress in the analysis

\item to speed up the computations by  making  use of an implicit
scheme for the solution of the equation \rf{3.1}. This will allow larger
time steps.

\end{itemize}

\Section{Appendix}
\bigskip
\noindent
{\large \bf Derivation of the 2-dimensional electrostatic equation}
\vspace{0.5cm}\\
\noindent
We consider a conducting 
strip made of a thin metal film, attached to a strip of non­zero conductivity substrate. 
The metal film may be continuous or it may be made of conducting patches with voids 
in between. We allow the metal film and substrate to have variable thickness. In 
the present formulation we neglect the interface resistance. 
The electrodes are attached to the strip and to the substrate.
We may want to compute the local field strength which 
determines the resulting electromigration. This is 
a more realistic model then the model based on the assumption of a zero conductivity 
substrate. It allows us also to consider the behaviour of a metal film with varying 
effective thickness at no extra cost.

\subsection{The 3-dimensional problem}

The 3-dimensional Problem Ohm's law implies: $\vec j = \sigma \vec E = 
-\sigma \nabla_3 \phi$, where $\vec j$ is the electric current density vector, 
$\vec E$ 
is the electric field vector, $\phi$ is the electric potential and 
$\sigma$ is the material conductivity. For steady fields Maxwell's equations 
with vanishing space charge give: 
\begin{equation}
\label{6.1}
\nabla_3 \cdot \vec j = 0,\ \mbox{where}\ \nabla_3=\left(\frac{\partial}{\partial x},
\frac{\partial}{\partial y},\frac{\partial}{\partial z}\right).
\end{equation}
Hence
\begin{equation}
\label{6.2}
\nabla_3 \cdot \left(\sigma \nabla_3 \phi\right) = 0.
\end{equation}
At all external (lateral) boundaries there is no flux in the direction of the normal,
$\vec n$, so that $\vec n \cdot \vec j = 0$, and using \rf{6.1} one gets:
\begin{equation}
\label{6.3}
\vec n \cdot \nabla_3 \phi = 0.
\end{equation}
%
The conditions \rf{6.3}, together with values of the potential specified at the 
strip and substrate edges and the continuity and jump conditions at the interface, constitute 
boundary conditions for equation \rf{6.2} in the two layers.
Thus the three 
dimensional potential can be found, in principle, as the solution of a well-posed 
three-dimensional boundary value problem. However such a solution can be very 
expensive to get in the present geometry, in particular as singularities in the 
solution will appear at sharp geometrical corners at crystal boundaries or voids, 
requiring high resolution or complicated integration formulae. To avoid this 
(probably unrealistic) behaviour of the solution and to avoid solving three dimensional 
problems many times, as required by the time development of the process, we 
proceed with an approximate approach suggested by (singular) perturbation theory. 

\subsection{The 2-dimensional equation}

We assume that $\phi$ and $\sigma$ change over a characteristic 
length scale $L$ in the horizontal directions $x$ and $y$ but over a scale $H$ 
in the vertical. 
Furthermore we assume that $\epsilon = H/L \ll 1$. 
Using scaled variables in \rf{6.2}, 
\begin{equation}
\label{6.4}
(X, Y, Z) = \left(x/L, y/L, z/H\right)
\end{equation}
we get:
\begin{equation}
\label{6.5}
\epsilon^2 \nabla_2(\sigma\nabla_2 \phi)+\frac{\partial}{\partial Z}\left(\sigma
\frac{\partial\phi}{\partial Z}\right)=0,\ \mbox{where}\ \nabla_2=\left(\frac{\partial}{\partial X},
\frac{\partial}{\partial Y}\right).
\end{equation}

Singular perturbation analysis considers an expansion
\begin{equation}
\label{6.6}
\phi=\phi_0+\epsilon^2\phi_1+\epsilon^4\phi_2+...
\end{equation}
where $\phi_k$ are functions of order $O(1)$ in $\epsilon$.
Substitution of \rf{6.6} in \rf{6.5} gives relations for the functions $\phi_k$ 
by grouping terms according to their order in $\epsilon$ and equating each group to zero. 
The zeroth order term gives : $\partial^2 \phi_0/\partial Z^2=0$,
thus $\phi_0$ is a linear function in $z$ for every $x$ 
and $y$ while taking into account \rf{6.3} kills off the $z$ dependence, so that: 
\begin{equation}
\label{6.7}
\phi_0=\phi_0(X,Y).
\end{equation}
Thus at this stage $\phi_0$ is an arbitrary function of the horizontal coordinates $X$
and $Y$. 
The first order equation and the boundary conditions in $Z$ result ultimately in 
the approximate two-dimensional equation for $\phi_0$ \cite{AIR}: 
\begin{equation}
\label{6.8}
\nabla_2(h_1\sigma_1+h_2\sigma_2)\nabla_2\phi_0=0
\end{equation}
where $h_1,\sigma_1$ and $h_2,\sigma_2$ are, respectively, the heights and conductivities of the two 
layers under consideration. The equation \rf{6.8} is solved with boundary conditions in 
the $(X,Y)$ plane. 
We remark that the approximate independence of the potential $\phi$ on the $Z$ 
coordinate justifies also the two dimensional approach for the electromigration equation. 
This behaviour is a consequence of the small aspect ratio assumption and the 
normal derivative boundary conditions \rf{6.3}, where one must also involve a small 
slope assumption.

\end{document}